\documentclass[showpacs,twocolumn,preprintnumbers,amsmath,amssymb]{revtex4} 
\pdfoutput=1 
\usepackage{amssymb}
\usepackage{mathrsfs}
\usepackage{graphicx}
\usepackage{dcolumn}
\usepackage{bm}
\usepackage{mathrsfs}
\usepackage{multirow}  
\usepackage{atlasphysics} 
\usepackage{lldefs} 
\usepackage{multirow}
\usepackage[hang,center]{subfigure} 
\usepackage[hyperindex,breaklinks]{hyperref}

\newcommand{\njets}{\ensuremath{N_{\textrm{jets}}}}

\newcommand{\bprime}{\ensuremath{b'}}                 
\newcommand{\md}{\ensuremath{m_{b'}}}                                                                 

\preprint{CERN-PH-EP-2011-230} 
\preprint{Submitted to Physical Review Letters}


\begin{document}

\title{
  Search for down-type fourth generation quarks with the ATLAS
  detector in events with one lepton and hadronically decaying $W$
  bosons
}

\author{The ATLAS Collaboration}

\begin{abstract}
  This Letter presents a search for pair production of heavy down-type
  quarks decaying via $\bprime \rightarrow Wt$ in the lepton + jets
  channel, as $\bprime\bar{\bprime}\rightarrow
  W^{-}tW^{+}\bar{t}\rightarrow
  b\bar{b}W^{+}W^{-}W^{+}W^{-}\rightarrow l^{\pm}\nu b\bar{b}
  q\bar{q}q\bar{q}q\bar{q}$. In addition to requiring exactly one
  lepton, large missing transverse momentum and at least six jets, the
  invariant mass of nearby jet pairs is used to identify high
  transverse momentum $W$ bosons. In data corresponding to an
  integrated luminosity of 1.04~\ifb{} from $pp$ collisions at
  $\sqrt{s}=7$ TeV recorded with the ATLAS detector, a heavy down-type
  quark with mass less than 480 \gev\ is excluded at the 95\%
  confidence level.
\end{abstract}

\maketitle

A fourth generation of chiral quarks is a natural extension to the
Standard Model (SM). It can explain some discrepancies observed
in meson-mixing data and can provide an additional source of CP violation in
$B_s$ decays.  A review of theoretical and experimental motivations
for a fourth generation of quarks can be found in Refs.~\cite{fourthgen,
  fourthgenconstraints}.

This Letter presents a search for a fourth generation down-type quark,
$b'$.  If $b'$ is chiral and its mass is larger than $m_t+m_W$, then
it decays predominantly as $b'\rightarrow Wt\rightarrow WWb$.  Pair
production of $b'$ quarks leads therefore to four $W$ bosons and two
$b$ quarks in the final state. This analysis applies more broadly to
any heavy quarks that decay into a $W$ boson and a $t$ quark, though
the fourth-generation $b'$ model is chosen as the benchmark.  The
previous limit in the single lepton channel is $\md>372$ \gev\ from
CDF, based on 4.8 \ifb{} of data~\cite{cdfbplj}. Searches using two or
more high \pt\ leptons in the final state have also been done at the
Tevatron~\cite{cdfbpss} and at the Large Hadron Collider
(LHC)~\cite{cmsbpss, atlasbpss, atlasss} with comparable sensitivity.

In the decay mode studied here, one of the four $W$ bosons decays
leptonically and the others decay hadronically. This lepton + jets
channel has more SM background than the mode with two $W$ bosons
decaying leptonically, but significantly larger acceptance. If the
mass difference between the $b'$ quark and the top quark is large, the
momentum of the $W$ boson from the $b'\rightarrow Wt$ decay is also
large, and the $W$ boson decay products become collimated.  At the
mass scales relevant to this search, the two quarks from the hadronic
$W$ decay give rise to two jets close to each other but still
resolvable in the detector as separate jets. The angle between the
decay products is related to the transverse momentum (\pt ) of the $W$
boson by $\Delta R \approx 2m_{W} / \pt^W $~\cite{coord}. To
distinguish the $b'$ signature from the SM backgrounds, the number of
jet pairs with small opening angle and invariant mass close to the $W$
boson mass is therefore used.

The major challenge for the lepton+jets mode is the estimation of the
SM background. The dominant source is \ttbar{} production with
additional jets, while $W+$ jets is the next most important
contribution. The significant theoretical uncertainty in the level of
gluon radiation affects the prediction of these backgrounds. As the
signal is distinguished from the background largely by the kinematic
properties of the jets, there also are significant experimental
uncertainties due to the energy scale and resolution of the jet energy
measurements.  Most of these uncertainties can be reduced by examining
signal-depleted samples which are sensitive to them. Other backgrounds
include single top, $Z$+jets where a lepton is not detected, and multijet
production in which a jet is misidentified as a lepton.

The data for this search were recorded with the \mbox{ATLAS}
detector~\cite{atlasdet}.  The momenta of charged particles with
pseudorapidity $|\eta| <$ 2.5 are measured with the inner detector
(ID), which includes a silicon pixel detector, a silicon microstrip
detector and a straw-tube detector, all operating in a uniform 2~T
axial magnetic field.  Electromagnetic (EM) calorimetry is provided by
a high-granularity, three-layer-depth sampling liquid argon detector
in the region $|\eta| <$ 3.2.  Jet reconstruction also uses hadronic
calorimetry provided by a scintillating tile detector with iron
absorbers in the region $|\eta| <$ 1.7, and liquid argon detectors
over \mbox{1.5 $< |\eta| <$ 4.9}. The muon spectrometer (MS) includes
tracking chambers for precision measurement in the bending plane up to
$|\eta| =$ 2.7 and fast trigger chambers up to $|\eta| =$ 2.4.  The
trigger chambers measure also the coordinate in the non-bending plane.
The muon detectors operate in a magnetic field generated by three
superconducting air-core toroids.

The events used in this analysis were selected using inclusive single
electron and muon triggers~\cite{trigger}. Electron candidates are
identified by localized energy deposits in the EM calorimeter with
transverse energy $E_{\rm T}>$ 20 \gev\ and $|\eta| <$ 2.47. The
energy cluster must satisfy shower-shape
requirements~\cite{ElectronPerformance} and should be matched with a
track reconstructed in the ID. Muon candidates must have transverse
momentum $p_{\rm T}>$ 18 \gev, $|\eta| <$ 2.4 and a consistent
trajectory reconstructed by combining segments in the ID and MS.

The data used in this search were collected in the first half of 2011,
and correspond to a total integrated luminosity of $1.04 \pm 0.04$
\ifb. During this period, the average number of collisions per bunch
crossing was six. The event reconstruction is affected by collisions
during the same bunch crossing as the selected event (in-time pileup)
and, to a lesser extent, collisions during adjacent bunch crossings,
within the time the detectors are sensitive for each trigger
(out-of-time pileup). The simulation takes both kinds of pileup into
account.

The signal and SM backgrounds are modeled using a variety of
generators. Pair-production of \bprime\ quarks decaying to $Wt$ with
subsequent showering and hadronization is generated with {\sc
  pythia}~\cite{pythia} using the MRST2007 LO* parton distribution
function (PDF) set~\cite{Sherstnev:2007nd}. Seven samples with $\md$
masses ranging from 300 to 600 \gev\ are used. The cross section for
each \bprime\ mass is calculated at approximate
next-to-next-to-leading order (NNLO) using Hathor~\cite{d4xs}. For a
\bprime\ quark with a mass of 350 \gev, the cross section is
$3.20^{+0.10}_{-0.19}{}^{+0.12}_{-0.12}$ pb, where the first
uncertainty comes from varying the renormalization and factorization
scales by a factor of two, and the second one from the PDFs. For a 500
\gev\ \bprime\ quark, the cross section is $0.33^{+
  0.01}_{-0.02}{}^{+0.01}_{-0.01}$ pb.

Top quark pair production is modeled using {\sc alpgen} \cite{alpgen}
where hard emission of up to three partons is described using QCD
matrix elements, {\sc herwig} \cite{herwig} is used to model the
parton shower, and {\sc jimmy}~\cite{jimmy} describes multiple parton
interactions. The rate of top quark production predicted by the
simulation is validated with data using an event sample with three, four
or five jets, where little or no $b'$ signal is expected.

Production of a $W$ or $Z$ boson in association with many jets is
described in {\sc alpgen} with hard parton emission of up to five
partons and {\sc herwig} for the parton shower. The $W$+jets
background is normalized using a data-driven method which fits
templates from simulated events to a data sample dominated by $W$ decays~\cite{topljets}.
The $Z$+jets background is normalized to a NNLO
calculation~\cite{fewz}.

Other processes considered are the production of dibosons ($WW, WZ,
ZZ$), modeled with {\sc alpgen} and {\sc herwig} and normalized to
next-to-leading-order (NLO) calculations~\cite{mcfm}; single top,
modeled with {\sc mc@nlo}~\cite{mcatnlo} and {\sc herwig}; and $\ttbar
W, \ttbar Z, \ttbar WW, \ttbar Wj, \ttbar Zj,$ and $WWjj$, all modeled
with {\sc madgraph}~\cite{madgraph} and {\sc pythia}.

The multijet background is strongly suppressed by the requirements
described below. The residual contribution is estimated using a
data-driven technique called the matrix method, described in detail in
Ref.~\cite{ljetspaper}.  Validation of this background estimate is
done by reversing these requirements to enhance the multijet
contribution.

Electrons, jets, muons, and missing transverse momentum are used to
select events for this search. Electrons are required to have $E_{\rm
  T} > 25$ \gev\ and be within the pseudorapidity range $|\eta| <
2.47$, excluding the barrel-endcap transition region $1.37 < |\eta| <
1.52$. Electrons must pass tight identification
requirements~\cite{ElectronPerformance} and also satisfy calorimeter
isolation: the energy not associated to the electron cluster inside a
cone of size $\Delta R = 0.2$ around the electron direction must be
smaller than $3.5$ \gev\, after the correction for the contributions
from interactions additional to the hard process.

Jets are reconstructed from topological calorimeter clusters using the
anti-$k_t$ algorithm~\cite{antiktalgorithm} with radius parameter 0.4.
These jets are then calibrated to the hadronic energy scale using \pt-
and $\eta$-dependent correction factors obtained from simulation and
validated with collision data~\cite{JetPerformance}.  For this
analysis, jets are required to satisfy $\pt > 25$ \gev\ and $|\eta| <
2.5$.  The closest jet within an $\eta$-$\phi$ cone of 0.2 around an
electron candidate is removed.

Muon candidates must satisfy \pt\ $> 20$ \gev\ and $|\eta| < 2.5$ and
pass tight identification requirements~\cite{ljetspaper}.  Muons must
also satisfy calorimeter isolation, which requires that the energy,
excluding the estimated energy deposited by the muon, is smaller than
4 \gev\ in a cone of size $\Delta R = 0.3$ around the muon direction,
and track isolation, which requires that the summed momentum of all
tracks excluding the muon track is smaller than 4 \gev\ in a cone of
size $\Delta R = 0.3$.  Finally, all muons within a cone of size
$\Delta R = 0.4$ around any jet with $\pt > 20$ \gev\ are removed.

The missing transverse momentum (\MET{}) is constructed from the
vector sum of topological calorimeter energy deposits and muon
momenta, projected onto the transverse plane~\cite{MET}.

If each $b'$ quark decays to a top quark and a $W$ boson, the
resulting final state is $\ttbar{}W^+W^-$.  In the lepton + jets decay
channel, the final state contains one lepton, \MET{} from the
undetected neutrino, and many jets from the eight quarks.  Exactly one
lepton ($e$ or $\mu$) must pass the selection described above. Since
not all jets are expected to satisfy the momentum and rapidity requirements, at
least six jets are required.

To reduce the multijet background, additional requirements are placed
on the \MET{} and the transverse mass of the leptonically decaying $W$
boson, $ m_{\mathrm T}^{W}=\sqrt{2 \MET
  \pt^{\ell}(1-\cos(\Delta\phi(\MET,\pt^{\ell})))}$.  In the electron
channel, \MET{}$>35$ \gev\ and ${m_{\mathrm T}^{W}}>25$ \gev\ are
required, and in the muon channel, \MET{}$>20$ \gev\ and
\MET{}$+m_{\mathrm T}^{W}>60$ \gev\ are required.  Only events with
six or more jets are considered. For a \bprime\ quark with a mass of
350 \gev, $11.2 \pm 1.7\%$ of signal events are accepted with this
selection. For a \bprime\ quark with a mass of 500 GeV, $13.5 \pm
2.0$\% of signal events are retained.

\begin{figure}[!htbp]
 \centering
 \includegraphics[width=0.37\textwidth]{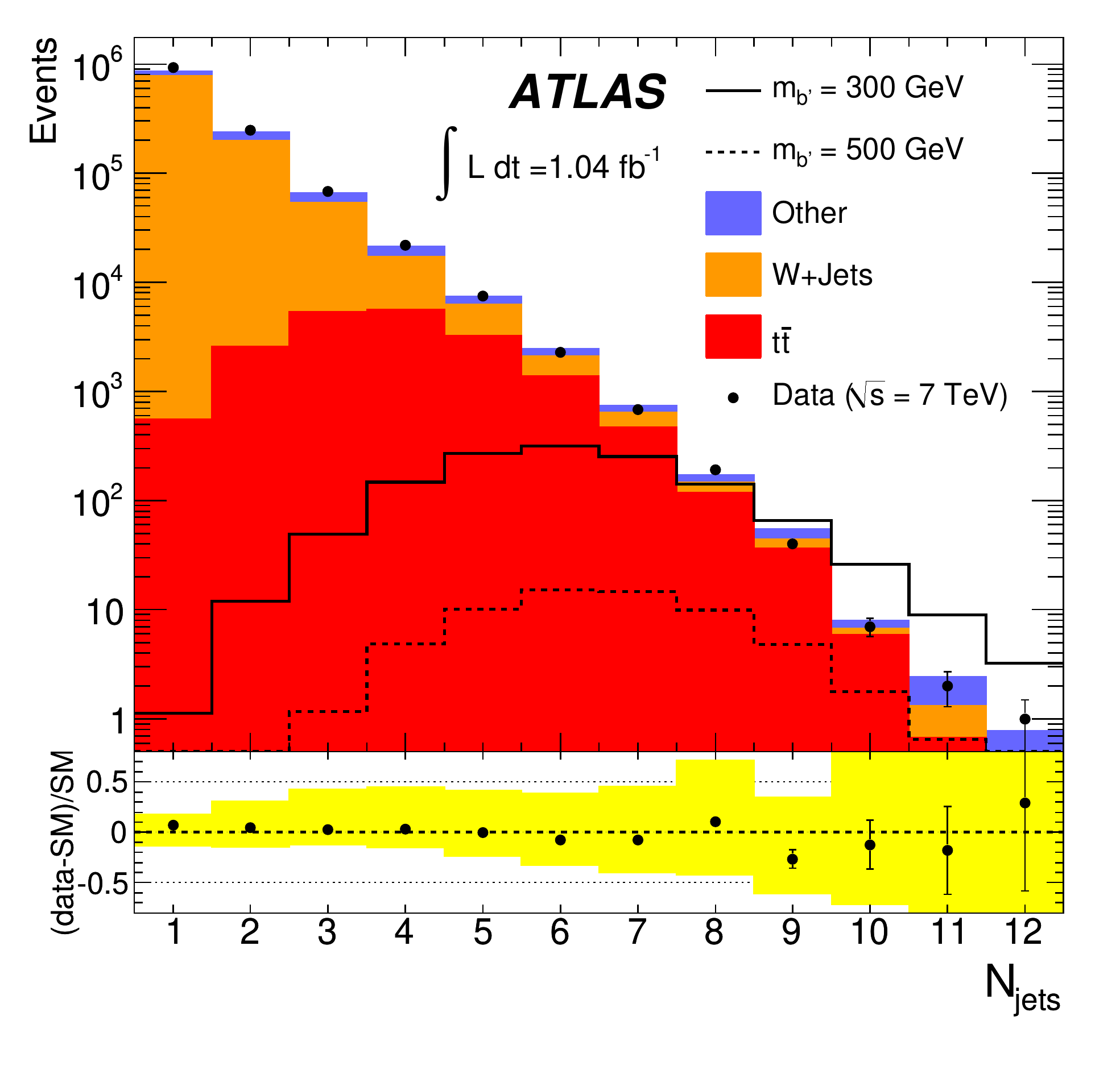}
 \caption{Jet multiplicity distribution for signal and backgrounds for
   events with at least one jet. The shaded SM backgrounds are stacked
   on one another, while the \bprime\ signal histograms are not. In
   this figure and those following, the bottom plot shows the relative
   difference between the SM prediction and the data together with the
   uncertainty (shaded band) due to statistics, jet energy scale, and $W$+jets
   normalization. }
 \label{fig:njets}
 \end{figure}

At this stage of the selection, pair production of
 \bprime\ quarks is distinguished mostly by the large number of
 energetic jets, as shown in Fig. ~\ref{fig:njets}.  Events with
 \bprime\ decays contain jets from three hadronic $W$ decays,
 while \ttbar{} background events contain only one hadronic $W$ decay.

 To identify these hadronic $W$ decays, pairs of jets separated by
 $\Delta R<1.0$ are examined. This choice of $\Delta R$ selects $W$
 bosons with high $\pt$ and reduces the combinatorial background in
 events with large jet multiplicity. The number of reconstructed $W$
 bosons ($N_W$) is defined as the number of such jet pairs with an
 invariant mass in the range $70-100$ \gev.  This range is not
 symmetric around the $W$ boson mass as additional energy is often
 included in the cone. Each jet may contribute to only one identified
 hadronic $W$ decay. In Fig.~\ref{fig:mjj}, the invariant masses of
 dijet pairs in a control sample of events with only three to five
 jets are shown. Good agreement is observed between the data and
 simulation across the entire spectrum including the region close to 
 the $W$ boson mass, where a bump can be seen in the \ttbar{} simulation.

\begin{figure}[!htbp]
\centering
\includegraphics[width=0.37\textwidth]{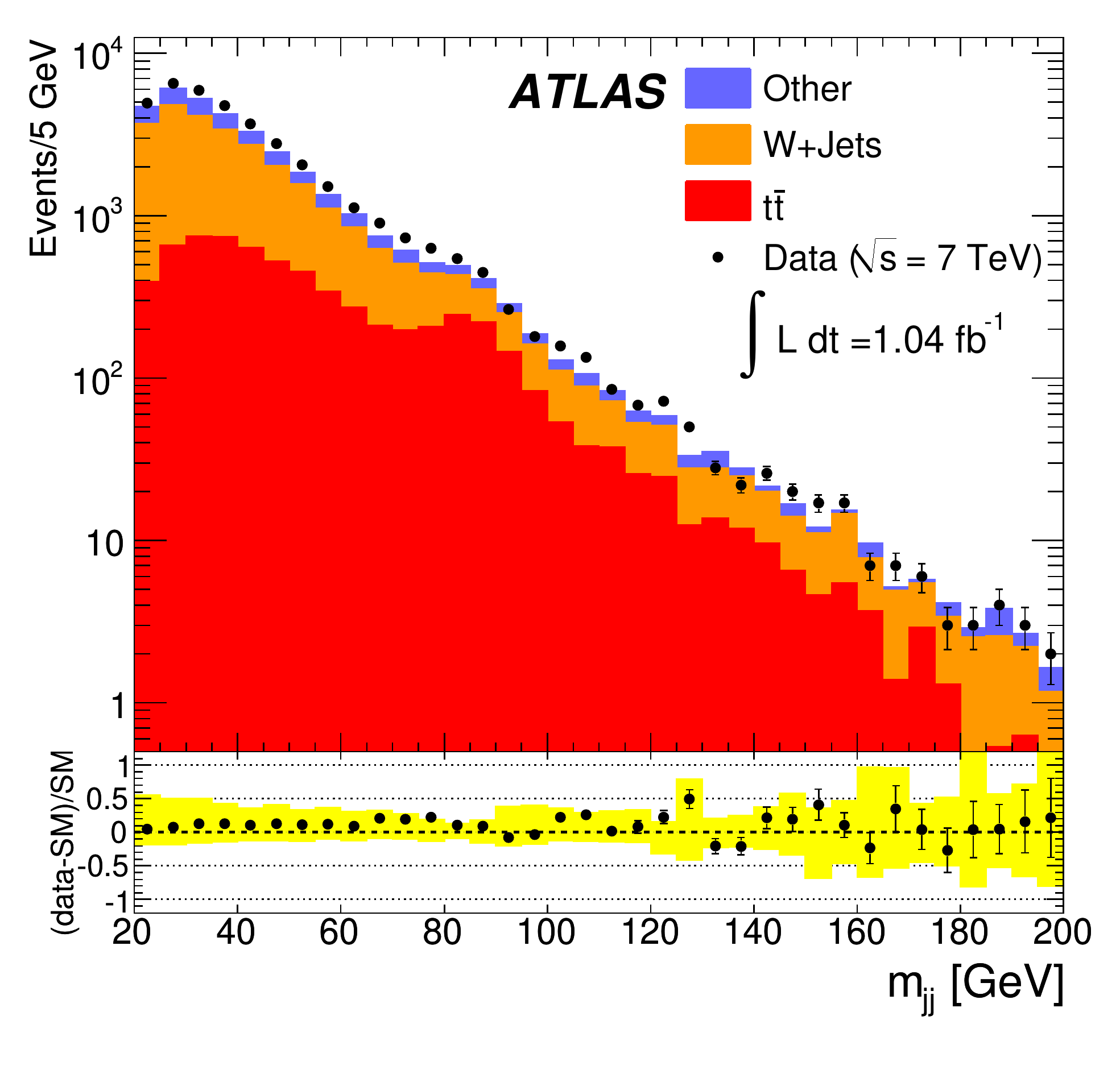}
\caption{The invariant mass distribution of jet pairs with $\Delta R<1.0$ for data and simulation in
  a control sample of events with exactly three to five jets.}
\label{fig:mjj}
\end{figure}

The efficiency of finding a simulated $W$ decay with both quarks
matched to separate reconstructed jets depends on the $W$ boson
\pt. For simulated \ttbar{} and $b'$ events passing the selection
described above and containing a $W$ boson with a \pt\ of about 250
\gev\, the two jets from the $W$ boson are found approximately 80\% of
the time. Once both jets are found, the efficiency that the jets have
$\Delta R<1.0$ and a dijet mass within the specified invariant mass range is approximately 70\%,
as can be seen in Fig.~\ref{fig:weff}.

\begin{figure}[!htbp]
\centering
\includegraphics[width=0.37\textwidth]{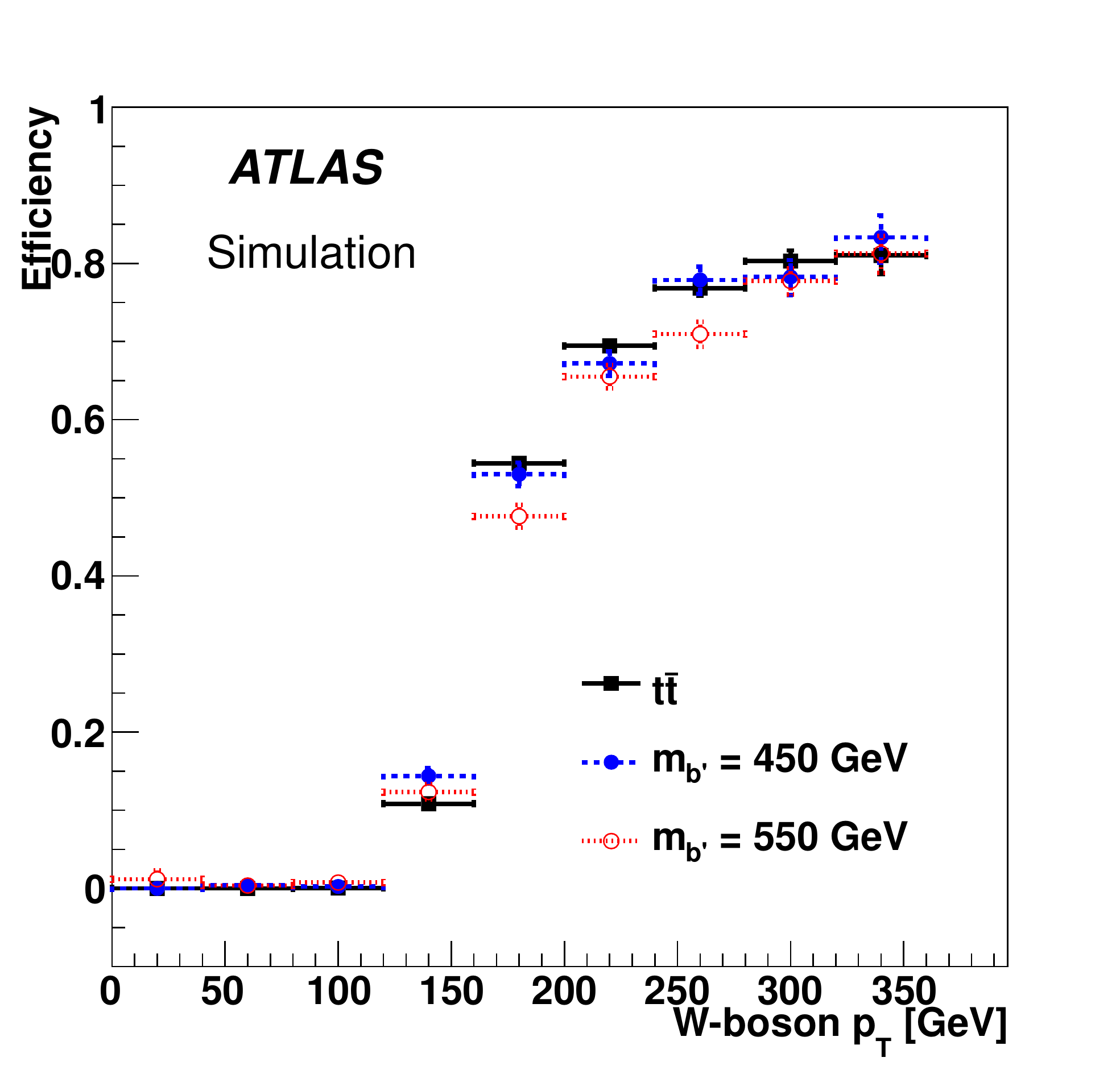}
\caption{The efficiency for jet pairs  from a
  simulated $W$ boson decay to have  $\Delta R<1.0$ and a dijet mass within $70-100$ GeV,
  for simulated \ttbar{} and signal events. Events are required to
  have exactly six jets.}
\label{fig:weff}
\end{figure}

To further distinguish the potential $b'$ signal from the backgrounds,
nine exclusive bins are examined, defined as a function of the
multiplicity of hadronic $W$ decays ($N_W=0,1,\ge 2$) and jet
multiplicity ($N_{\textrm{jet}}=6,7,\ge 8$). 

The agreement between data and simulation for the description of the
number of jets is validated in events with a scalar sum ($H_{\mathrm
  T}$) of transverse energies of jets and leptons less than 400 \gev\
and no reconstructed hadronic $W$ decays, to suppress potential $b'$
contributions.

\begin{table}
  \caption{Systematic uncertainties in the predicted total
    background in the signal region.  Some of the uncertainties have
    been constrained in background-dominated regions, `profiled' as
    described in the text.  Smaller systematic uncertainties, 
    such as those related to lepton identification and theory, 
    and small uncertainties on the rate, are not profiled and are 
    not included here. For the profiled systematics, the uncertainty 
    before profiling is given in parentheses.}
\label{tab:syst}

\begin{tabular}{lr}
\hline\hline
 & Uncertainty on Background \\ \hline
Profiled Uncertainties &\\ \hline
$W$+jets Normalization & $\pm5\%$\ ($\pm16\%$)\\
ISR/FSR & $\pm12\%\ $($\pm17\%$) \\
Jet Energy Resolution& $\pm3\%\ $($\pm6\%$)\\
Jet Reconstruction Efficiency& $\pm2\%\ $($\pm3\%$)\\ \hline
Not-profiled Uncertainties &\\ \hline
Jet Energy Scale &$\pm31\%$\\ 
\ttbar{} Simulation Generator &$\pm6\%$\\ 
\ttbar{} Showering Model& $\pm3\%$\\ 
\hline\hline
\end{tabular}
\end{table}

Table~\ref{tab:syst} shows the major sources of systematic
uncertainty. The main contributions to uncertainty in the modeling of
the backgrounds and $b'$ signal come from the jet energy scale and the
level of initial and final state radiation (ISR/FSR) in the top quark
pair background. The jet energy scale uncertainty is extracted from
dijet events and validated with $\gamma+$jet events as discussed in
Ref.~\cite{JetPerformance}, with an additional uncertainty due to
in-time pileup. The amounts of simulated ISR and FSR are varied
according to their uncertainties for both background and signal
events. Jet reconstruction efficiency and jet energy resolution lead
to smaller uncertainties in the predicted background.

 For the largest background source,
\ttbar{} with additional jets, uncertainties in the description of the
parton shower and fragmentation model are estimated by comparing
predictions of {\sc powheg}~\cite{powheg} with {\sc pythia} to {\sc
powheg} with {\sc herwig}. Uncertainties in the modeling of the
production and decay of the top quark are estimated by comparing the
predictions from {\sc powheg} with {\sc herwig} and {\sc alpgen}.

The $W$+jets normalization uncertainty is $4\%$, plus $24\%$ per jet
added in quadrature~\cite{topljets}. The uncertainties in lepton
reconstruction efficiency and energy scale are derived in dilepton
samples dominated by $Z\rightarrow \ell\ell$ decays and applied to the
simulated background and signal samples.

The systematic uncertainties are treated as correlated between signal
and background, and between electron and muon channels, except where
they are specific to the background model (e.g. $W+$jets
normalization) or to a channel (e.g. electron or muon efficiencies).

To extract the most likely value of the \antibar{\bprime} cross section
in the nine bins of ($N_W,N_{\textrm{jet}}$) multiplicity, a binned
maximum likelihood fit using a profile likelihood ratio is performed,
varying each background rate within its uncertainty, and allowing shape and
rate variation due to the systematic uncertainties described above.  The
signal and background rates are fitted simultaneously.

Events in the final selection which have low hadronic $W$ boson or jet multiplicity ($N_W < 2$ and $N_{\rm
  jet} <$ 8) are dominated by
background processes and serve to constrain some of the systematic
uncertainties. The likelihood is maximized with respect to the
variation due to the systematic uncertainties. This procedure serves
to reduce some of the systematic uncertainties, those listed as 'profiled' in Table~\ref{tab:syst}.

The expected background and signal contributions, as well as the
observed numbers of events in the data, are shown in
Fig.~\ref{fig:sig_bins} and given in Table~\ref{tab:bg} for the nine
bins of jet and hadronic $W$-boson multiplicity. No evidence
for the production of $b'$ quarks is observed. The CLs
method~\cite{MCLimit} is used to set 95\% confidence level (C.L.) cross section upper limits
for the pair production of fourth generation quarks, $b'$.  The median
expected upper limit is extracted in the background-only hypothesis.
The results are shown in Fig.~\ref{fig:exp_L1035_syst} as a function
of the $b'$ mass. Systematic uncertainties are taken into account and
it is assumed that the branching ratio (BR) for $b' \rightarrow Wt$ is
100 \%. These cross section limits are interpreted as limits on the
$b'$ mass by finding the intersection of the limit curves with the
theoretical cross section curve.  Uncertainty in the theoretical cross
section includes renormalization and factorization scale and PDF
uncertainties calculated with Hathor~\cite{d4xs}.

Masses below 480 \gev\ are excluded at the 95\% confidence level, while
the expected limit is $m_{b'} >$ 470 \gev. For a particle with
a mass of 480 \gev, the expected exclusion limit on the pair
production cross section is $\sigma < 0.54^{+0.45}_{-0.22}$ pb, while
the observed exclusion is $\sigma < 0.47$ pb.

\begin{figure}[!htbp]
\centering
\includegraphics[width=0.42\textwidth]{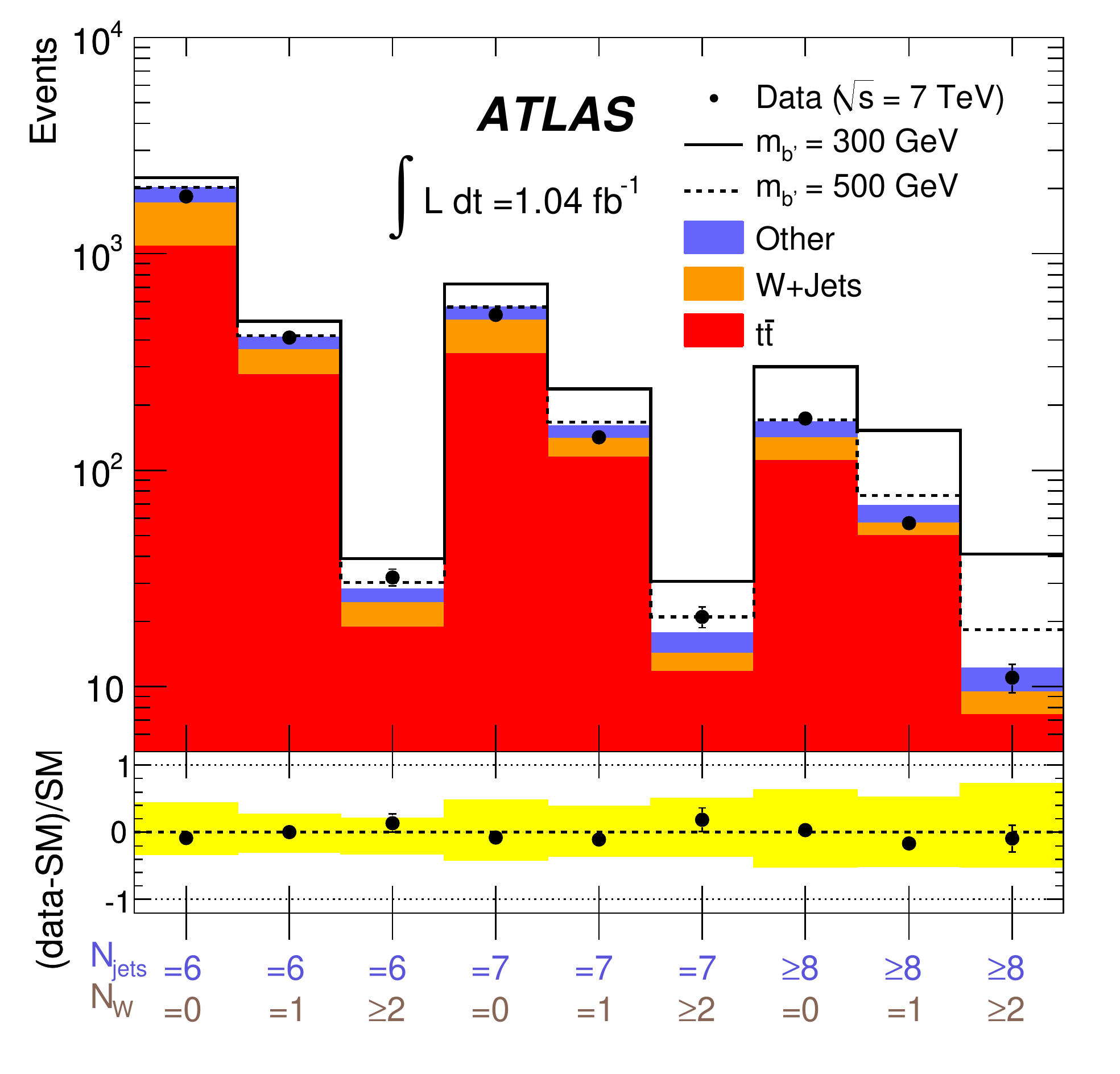}
\caption{ Distribution of the numbers of events observed in the data
  and expected from SM processes for jet multiplicity $\njets = 6$,
  $7$, $\geq8$ with hadronic $W$ multiplicity 
  \mbox{$N_{W} =0,1,\geq2$}. The expected b' signals for two masses are also
  shown, stacked on top of the backgrounds.}
\label{fig:sig_bins}
\end{figure}

\begin{table}
  \caption{Expected and observed number of events in each bin of jet and hadronic $W$ decay
    multiplicity. Estimates for two signal samples with different $b'$ masses are also shown. The contributions from
    different background sources are shown in Fig.~\ref{fig:sig_bins}.}
\label{tab:bg}

 \begin{tabular}{llrrrrrr}
 \hline\hline
   &  & Expected  & Observed & $b'$ 350 & $b'$ 500 \\
 $N_{\mathrm{jet}}$& $N_{W}$ & background &  events &   \gev\ &
 \gev\ \\ \hline
 \vspace{1mm} 6&0 & $2060^{+850}_{-750}$ &  1839 & 80 & 5 \\
 \vspace{1mm}6&1 & $410^{+104}_{-150}$ &  410 & 47 & 8 \\
 6&$\ge 2$ &  $28^{+10}_{-16}$ &  32 & 7 & 2  \\
 \vspace{1mm}7&0 & $570^{+320}_{-230}$ &  521 & 60 & 4  \\
 \vspace{1mm}7&1 & $166^{+49}_{-68}$ &  142 & 46 & 7  \\
 \vspace{1mm}7&$\ge 2$ & $17.9^{+6.6}_{-6.8}$ &  21 & 11 & 3  \\
 \vspace{1mm}$\ge 8$&0 & $170^{+180}_{-70}$ &  173 & 56 & 3  \\
 \vspace{1mm}$\ge 8$&1 & $69^{+33}_{-27}$ &  57 & 50 & 8  \\
 \vspace{1mm}$\ge 8$&$\ge 2$ & $12.1^{+8.6}_{-5.2}$ &  11 & 22 & 6  \\
 \hline\hline
 \end{tabular}
 \end{table}

\begin{figure}[!htbp]
\centering
\includegraphics[width=0.4\textwidth]{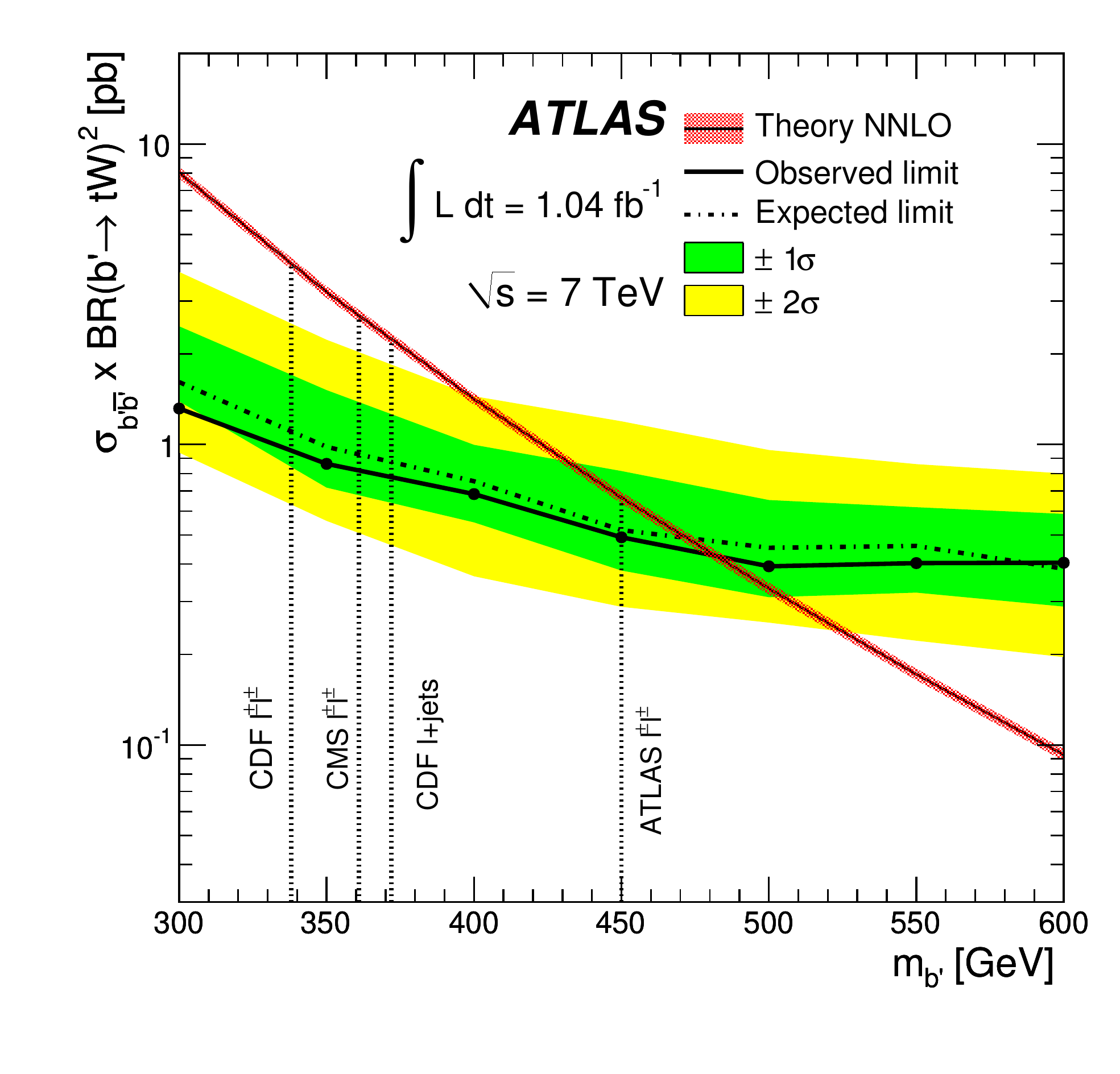}
\caption{ Expected and observed cross section exclusion upper limits
  at $95\%$ C.L. for a fourth-generation $b'$ quark. Systematic
  uncertainties on the expected limit are shown with shaded
  bands. Previously published limits from CDF~\cite{cdfbpss,cdfbplj}, CMS~\cite{cmsbpss}, and ATLAS~\cite{atlasss} are also shown.}
\label{fig:exp_L1035_syst}
\end{figure}

In conclusion, a search for pair production of heavy down-type quarks
decaying via $\bprime\rightarrow Wt$ in the lepton + jets channel has
been performed using 1.04~\ifb{} of $\sqrt{s} = 7$ TeV $pp$
collision data recorded with the ATLAS detector, selecting events
based on the number of jets and hadronic $W$ decays. A heavy down-type
quark with mass less than 480 \gev\ is excluded at the 95\%
confidence level, improving significantly on previous limits.


\section{Acknowledgements}

We thank CERN for the very successful operation of the LHC, as well as
the support staff from our institutions without whom ATLAS could not
be operated efficiently.

We acknowledge the support of ANPCyT, Argentina; YerPhI, Armenia; ARC,
Australia; BMWF, Austria; ANAS, Azerbaijan; SSTC, Belarus; CNPq and
FAPESP, Brazil; NSERC, NRC and CFI, Canada; CERN; CONICYT, Chile; CAS,
MOST and NSFC, China; COLCIENCIAS, Colombia; MSMT CR, MPO CR and VSC
CR, Czech Republic; DNRF, DNSRC and Lundbeck Foundation, Denmark;
EPLANET and ERC, European Union; IN2P3-CNRS, CEA-DSM/IRFU, France;
GNAS, Georgia; BMBF, DFG, HGF, MPG and AvH Foundation, Germany; GSRT,
Greece; ISF, MINERVA, GIF, DIP and Benoziyo Center, Israel; INFN,
Italy; MEXT and JSPS, Japan; CNRST, Morocco; FOM and NWO, Netherlands;
RCN, Norway; MNiSW, Poland; GRICES and FCT, Portugal; MERYS (MECTS),
Romania; MES of Russia and ROSATOM, Russian Federation; JINR; MSTD,
Serbia; MSSR, Slovakia; ARRS and MVZT, Slovenia; DST/NRF, South
Africa; MICINN, Spain; SRC and Wallenberg Foundation, Sweden; SER,
SNSF and Cantons of Bern and Geneva, Switzerland; NSC, Taiwan; TAEK,
Turkey; STFC, the Royal Society and Leverhulme Trust, United Kingdom;
DOE and NSF, United States of America.

The crucial computing support from all WLCG partners is acknowledged
gratefully, in particular from CERN and the ATLAS Tier-1 facilities at
TRIUMF (Canada), NDGF (Denmark, Norway, Sweden), CC-IN2P3 (France),
KIT/GridKA (Germany), INFN-CNAF (Italy), NL-T1 (Netherlands), PIC
(Spain), ASGC (Taiwan), RAL (UK) and BNL (USA) and in the Tier-2
facilities worldwide.


\onecolumngrid

\clearpage
\begin{flushleft}
{\Large The ATLAS Collaboration}

\bigskip

G.~Aad$^{\rm 48}$,
B.~Abbott$^{\rm 110}$,
J.~Abdallah$^{\rm 11}$,
A.A.~Abdelalim$^{\rm 49}$,
A.~Abdesselam$^{\rm 117}$,
O.~Abdinov$^{\rm 10}$,
B.~Abi$^{\rm 111}$,
M.~Abolins$^{\rm 87}$,
O.S.~AbouZeid$^{\rm 157}$,
H.~Abramowicz$^{\rm 152}$
% ...
\bigskip

% ...
$^{10}$ Institute of Physics, Azerbaijan Academy of Sciences, Baku, Azerbaijan\\
$^{11}$ Institut de F\'isica d'Altes Energies and Departament de F\'isica de la Universitat Aut\`onoma  de Barcelona and ICREA, Barcelona, Spain\\
% ...
$^{48}$ Fakult\"{a}t f\"{u}r Mathematik und Physik, Albert-Ludwigs-Universit\"{a}t, Freiburg i.Br., Germany\\
$^{49}$ Section de Physique, Universit\'e de Gen\`eve, Geneva, Switzerland\\
% ...
$^{87}$ Department of Physics and Astronomy, Michigan State University, East Lansing MI, United States of America\\
% ...
$^{110}$ Homer L. Dodge Department of Physics and Astronomy, University of Oklahoma, Norman OK, United States of America\\
$^{111}$ Department of Physics, Oklahoma State University, Stillwater OK, United States of America\\
% ...
$^{117}$ Department of Physics, Oxford University, Oxford, United Kingdom\\
% ...
$^{152}$ Raymond and Beverly Sackler School of Physics and Astronomy, Tel Aviv University, Tel Aviv, Israel\\
% ...
$^{157}$ Department of Physics, University of Toronto, Toronto ON, Canada\\
% ...
\end{flushleft}



\begin{thebibliography}{99}
\bibitem{coord} ATLAS uses a right-handed coordinate system with its
  origin at the nominal interaction point (IP) in the center of the
  detector and the $z$-axis along the beam pipe. The $x$-axis points
  from the IP to the center of the LHC ring; the $y$-axis points
  upward. Cylindrical coordinates $(r,\phi)$ are used in the
  transverse plane, $\phi$ being the azimuthal angle around the beam
  pipe. The pseudorapidity is defined in terms of the polar angle
  $\theta$ as $\eta=-\ln\tan(\theta/2)$. A cone in $\eta$-$\phi$
  is defined as $\Delta R$ is defined as $\sqrt{(\Delta \eta)^2 + (\Delta
    \phi)^2}$.
\end{thebibliography}
\end{document}